\documentclass[twocolumn,aps,tightenlinefs,superscriptaddress,preprintnumbers]{revtex4}
\usepackage{graphicx,color,dcolumn,booktabs,bm}
\usepackage{longtable,lscape}
\usepackage{amsmath}
\usepackage{indentfirst}
\usepackage{epsfig}
\usepackage{feynmf}
\usepackage{epstopdf}
\usepackage{slashed}
\usepackage{cases}
\usepackage{color}
\usepackage{multirow}
\usepackage{ulem}
\usepackage{graphicx,color,dcolumn,booktabs,bm}
\usepackage{verbatim}
\usepackage[colorlinks,linkcolor=red,anchorcolor=green,citecolor=blue]{hyperref}
\usepackage{mathrsfs}
\usepackage{rotating}
\usepackage{threeparttable}
\usepackage{dcolumn}

\newcommand{\BR}{{\cal B}}
\newcommand{\W}{{\cal W}}
\newcommand{\jp}{J^{P}}
\newcommand{\pip}{\pi^+}
\newcommand{\pim}{\pi^-}

\newcommand{\EE}{e^+e^-}

\newcommand{\beq}{\begin{equation}}
\newcommand{\eeq}{\end{equation}}
\newcommand{\bitm}{\begin{itemize}}
\newcommand{\eitm}{\end{itemize}}

\newcommand{\ones}{\Upsilon(1S)}
\newcommand{\twos}{\Upsilon(2S)}

\newcommand{\threes}{\Upsilon(3S)}
\newcommand{\fours}{\Upsilon(4S)}
\newcommand{\fives}{\Upsilon(5S)}

\begin{document}
\hyphenpenalty=10000


\title{\quad\\[0.1cm]\boldmath Evidence for the decay $\Omega_{c}^{0} \to \pi^+\Omega(2012)^- \to \pi^+ (\bar{K}\Xi)^{-}$}


\noaffiliation
\affiliation{Department of Physics, University of the Basque Country UPV/EHU, 48080 Bilbao}
\affiliation{University of Bonn, 53115 Bonn}
\affiliation{Brookhaven National Laboratory, Upton, New York 11973}
\affiliation{Budker Institute of Nuclear Physics SB RAS, Novosibirsk 630090}
\affiliation{Faculty of Mathematics and Physics, Charles University, 121 16 Prague}
\affiliation{Chonnam National University, Gwangju 61186}
\affiliation{University of Cincinnati, Cincinnati, Ohio 45221}
\affiliation{Deutsches Elektronen--Synchrotron, 22607 Hamburg}
\affiliation{University of Florida, Gainesville, Florida 32611}
\affiliation{Department of Physics, Fu Jen Catholic University, Taipei 24205}
\affiliation{Key Laboratory of Nuclear Physics and Ion-beam Application (MOE) and Institute of Modern Physics, Fudan University, Shanghai 200443}
\affiliation{Gifu University, Gifu 501-1193}
\affiliation{SOKENDAI (The Graduate University for Advanced Studies), Hayama 240-0193}
\affiliation{Gyeongsang National University, Jinju 52828}
\affiliation{Department of Physics and Institute of Natural Sciences, Hanyang University, Seoul 04763}
\affiliation{University of Hawaii, Honolulu, Hawaii 96822}
\affiliation{High Energy Accelerator Research Organization (KEK), Tsukuba 305-0801}
\affiliation{J-PARC Branch, KEK Theory Center, High Energy Accelerator Research Organization (KEK), Tsukuba 305-0801}
\affiliation{National Research University Higher School of Economics, Moscow 101000}
\affiliation{Forschungszentrum J\"{u}lich, 52425 J\"{u}lich}
\affiliation{IKERBASQUE, Basque Foundation for Science, 48013 Bilbao}
\affiliation{Indian Institute of Technology Bhubaneswar, Satya Nagar 751007}
\affiliation{Indian Institute of Technology Hyderabad, Telangana 502285}
\affiliation{Indian Institute of Technology Madras, Chennai 600036}
\affiliation{Indiana University, Bloomington, Indiana 47408}
\affiliation{Institute of High Energy Physics, Chinese Academy of Sciences, Beijing 100049}
\affiliation{Institute of High Energy Physics, Vienna 1050}
\affiliation{Institute for High Energy Physics, Protvino 142281}
\affiliation{INFN - Sezione di Napoli, 80126 Napoli}
\affiliation{INFN - Sezione di Roma Tre, I-00146 Roma}
\affiliation{INFN - Sezione di Torino, 10125 Torino}
\affiliation{Advanced Science Research Center, Japan Atomic Energy Agency, Naka 319-1195}
\affiliation{J. Stefan Institute, 1000 Ljubljana}
\affiliation{Institut f\"ur Experimentelle Teilchenphysik, Karlsruher Institut f\"ur Technologie, 76131 Karlsruhe}
\affiliation{Department of Physics, Faculty of Science, King Abdulaziz University, Jeddah 21589}
\affiliation{Kitasato University, Sagamihara 252-0373}
\affiliation{Korea Institute of Science and Technology Information, Daejeon 34141}
\affiliation{Korea University, Seoul 02841}
\affiliation{Kyoto Sangyo University, Kyoto 603-8555}
\affiliation{Kyungpook National University, Daegu 41566}
\affiliation{Universit\'{e} Paris-Saclay, CNRS/IN2P3, IJCLab, 91405 Orsay}
\affiliation{P.N. Lebedev Physical Institute of the Russian Academy of Sciences, Moscow 119991}
\affiliation{Liaoning Normal University, Dalian 116029}
\affiliation{Faculty of Mathematics and Physics, University of Ljubljana, 1000 Ljubljana}
\affiliation{Ludwig Maximilians University, 80539 Munich}
\affiliation{Luther College, Decorah, Iowa 52101}
\affiliation{Malaviya National Institute of Technology Jaipur, Jaipur 302017}
\affiliation{Faculty of Chemistry and Chemical Engineering, University of Maribor, 2000 Maribor}
\affiliation{Max-Planck-Institut f\"ur Physik, 80805 M\"unchen}
\affiliation{University of Mississippi, University, Mississippi 38677}
\affiliation{University of Miyazaki, Miyazaki 889-2192}
\affiliation{Moscow Physical Engineering Institute, Moscow 115409}
\affiliation{Graduate School of Science, Nagoya University, Nagoya 464-8602}
\affiliation{Universit\`{a} di Napoli Federico II, 80126 Napoli}
\affiliation{Nara Women's University, Nara 630-8506}
\affiliation{National Central University, Chung-li 32054}
\affiliation{National United University, Miao Li 36003}
\affiliation{Department of Physics, National Taiwan University, Taipei 10617}
\affiliation{H. Niewodniczanski Institute of Nuclear Physics, Krakow 31-342}
\affiliation{Nippon Dental University, Niigata 951-8580}
\affiliation{Niigata University, Niigata 950-2181}
\affiliation{Novosibirsk State University, Novosibirsk 630090}
\affiliation{Osaka City University, Osaka 558-8585}
\affiliation{Pacific Northwest National Laboratory, Richland, Washington 99352}
\affiliation{Panjab University, Chandigarh 160014}
\affiliation{University of Pittsburgh, Pittsburgh, Pennsylvania 15260}
\affiliation{Research Center for Nuclear Physics, Osaka University, Osaka 567-0047}
\affiliation{Meson Science Laboratory, Cluster for Pioneering Research, RIKEN, Saitama 351-0198}
\affiliation{Department of Modern Physics and State Key Laboratory of Particle Detection and Electronics, University of Science and Technology of China, Hefei 230026}
\affiliation{Seoul National University, Seoul 08826}
\affiliation{Showa Pharmaceutical University, Tokyo 194-8543}
\affiliation{Soongsil University, Seoul 06978}
\affiliation{Sungkyunkwan University, Suwon 16419}
\affiliation{School of Physics, University of Sydney, New South Wales 2006}
\affiliation{Department of Physics, Faculty of Science, University of Tabuk, Tabuk 71451}
\affiliation{Tata Institute of Fundamental Research, Mumbai 400005}
\affiliation{Department of Physics, Technische Universit\"at M\"unchen, 85748 Garching}
\affiliation{School of Physics and Astronomy, Tel Aviv University, Tel Aviv 69978}
\affiliation{Department of Physics, Tohoku University, Sendai 980-8578}
\affiliation{Earthquake Research Institute, University of Tokyo, Tokyo 113-0032}
\affiliation{Department of Physics, University of Tokyo, Tokyo 113-0033}
\affiliation{Tokyo Institute of Technology, Tokyo 152-8550}
\affiliation{Tokyo Metropolitan University, Tokyo 192-0397}
\affiliation{Virginia Polytechnic Institute and State University, Blacksburg, Virginia 24061}
\affiliation{Wayne State University, Detroit, Michigan 48202}
\affiliation{Yamagata University, Yamagata 990-8560}
\affiliation{Yonsei University, Seoul 03722}

\author{Y.~Li}\affiliation{Key Laboratory of Nuclear Physics and Ion-beam Application (MOE) and Institute of Modern Physics, Fudan University, Shanghai 200443} 
\author{S.~S.~Tang}\affiliation{Key Laboratory of Nuclear Physics and Ion-beam Application (MOE) and Institute of Modern Physics, Fudan University, Shanghai 200443} 
\author{S.~Jia}\affiliation{Key Laboratory of Nuclear Physics and Ion-beam Application (MOE) and Institute of Modern Physics, Fudan University, Shanghai 200443} 
\author{C.~P.~Shen}\affiliation{Key Laboratory of Nuclear Physics and Ion-beam Application (MOE) and Institute of Modern Physics, Fudan University, Shanghai 200443} 
\author{I.~Adachi}\affiliation{High Energy Accelerator Research Organization (KEK), Tsukuba 305-0801}\affiliation{SOKENDAI (The Graduate University for Advanced Studies), Hayama 240-0193} 
\author{H.~Aihara}\affiliation{Department of Physics, University of Tokyo, Tokyo 113-0033} 
\author{S.~Al~Said}\affiliation{Department of Physics, Faculty of Science, University of Tabuk, Tabuk 71451}\affiliation{Department of Physics, Faculty of Science, King Abdulaziz University, Jeddah 21589} 
\author{D.~M.~Asner}\affiliation{Brookhaven National Laboratory, Upton, New York 11973} 
\author{H.~Atmacan}\affiliation{University of Cincinnati, Cincinnati, Ohio 45221} 
\author{V.~Aulchenko}\affiliation{Budker Institute of Nuclear Physics SB RAS, Novosibirsk 630090}\affiliation{Novosibirsk State University, Novosibirsk 630090} 
% ...
\collaboration{The Belle Collaboration}

\begin{abstract}
Using a data sample of 980~fb$^{-1}$ collected with the Belle detector operating at the
KEKB asymmetric-energy $e^+e^-$ collider, we present evidence for the $\Omega(2012)^-$ in
the resonant substructure of $\Omega_{c}^{0} \to \pi^+ (\bar{K}\Xi)^{-}$
($(\bar{K}\Xi)^{-}$ = $K^-\Xi^0$ + $\bar{K}^0 \Xi^-$) decays. The significance of the
$\Omega(2012)^-$ signal is 4.2$\sigma$ after considering the systematic uncertainties.
The ratio of the branching fraction of $\Omega_{c}^{0} \to \pi^{+}
\Omega(2012)^- \to \pi^+ (\bar{K}\Xi)^{-}$ relative to that of $\Omega_{c}^{0} \to \pi^{+}
\Omega^-$ is calculated to be 0.220 $\pm$ 0.059(stat.) $\pm$ 0.035(syst.).
The individual ratios of the branching fractions of the two isospin modes
are also determined, and found to be ${\cal B}(\Omega_{c}^0 \to \pi^+ \Omega(2012)^-) \times {\cal B}(\Omega(2012)^-
\to K^-\Xi^0)/{\cal B}(\Omega_{c}^0 \to \pi^+  K^- \Xi^0)$ = (9.6 $\pm$ 3.2(stat.) $\pm$ 1.8(syst.))\%
and ${\cal B}(\Omega_{c}^0 \to \pi^+ \Omega(2012)^-) \times {\cal B}(\Omega(2012)^- \to \bar{K}^0 \Xi^-)/{\cal B}(\Omega_{c}^0
\to \pi^+  \bar{K}^0 \Xi^-)$ = (5.5 $\pm$ 2.8(stat.) $\pm$ 0.7(syst.))\%.
\end{abstract}


\maketitle
Several excited $\Omega^-$ baryons have been observed~\cite{PDG};
the latest addition was an excited $\Omega^{-}$ state decaying into $K^-\Xi^0$ and
$K_S^0 \Xi^-$ observed by Belle in 2018 using data samples collected at the $\Upsilon(1S)$,
$\Upsilon(2S)$, and $\Upsilon(3S)$ resonances~\cite{Omega}. This new excited
$\Omega^-$ state is called the $\Omega(2012)^-$, and has a measured mass of
(2012.4 $\pm$ 0.7(stat.) $\pm$ 0.6(syst.))~MeV/$c^2$ and width of
(6.4$^{+2.5}_{-2.0}$(stat.) $\pm$ 1.6(syst.))~MeV.

Following the discovery of the $\Omega(2012)^-$, several interpretations of the state were
suggested~\cite{theory1,theory2,theory3,theory4,theory5,theory6,theory7}.
The mass and the two-body strong decays of the $\Omega(2012)^-$ were studied in the
framework of Quantum Chromodynamics sum rules~\cite{theory1,theory2}, and this showed that
the $\Omega(2012)^-$ could be interpreted as a 1$P$ orbital excitation of the ground-state
$\Omega^-$ baryon with a spin-parity $\jp = 3/2^-$. As the mass of the $\Omega(2012)^-$
is very close to the $(\bar{K}\Xi(1530))^-$ threshold, it was interpreted as a
$(\bar{K}\Xi(1530))^-$ hadronic molecule in Refs.~\cite{theory3,theory4,theory5,theory6,theory7}.
These hadronic molecule models predicted a large decay width for $\Omega(2012)^- \to (\bar{K}\pi\Xi)^-$.

The three-body decay $\Omega(2012)^- \to (\bar{K}\Xi(1530))^- \to (\bar{K} \pi \Xi)^-$ has been
searched for by Belle~\cite{Jia:2019eav}. No significant signals were found for the
$\Omega(2012)^- \to (\bar{K}\Xi(1530))^- \to (\bar{K} \pi \Xi)^-$ decay, and the 90\%
credibility level (C.L.) upper limit on the ratio of ${\cal R}_{(\bar{K}\Xi)^-}^{(\bar{K}\pi\Xi)^-}$ =
$\BR(\Omega(2012)^- \to (\bar{K}\Xi(1530))^- \to (\bar{K} \pi \Xi)^-)$/
$\BR(\Omega(2012)^- \to (\bar{K} \Xi)^-)$ was determined to be 0.119.
Based on this upper limit for the ratio ${\cal R}_{(\bar{K}\Xi)^-}^{(\bar{K}\pi\Xi)^-}$,
the authors in Refs.~\cite{Ikeno:2020vqv,Lu:2020ste} revisited the $\Omega(2012)^-$
resonance from the molecular perspective, and concluded that the experimental
data was still consistent with their molecular picture with a certain set of naturally allowed
parameters. On the other hand, the authors of Ref.~\cite{Liu:2020yen}
conducted a dynamical calculation of pentaquark systems with quark contents $sssu\bar{u}$ in the
framework of the chiral quark model~\cite{Valcarce:2005em} and the quark delocalization color
screening model~\cite{Wang:1992wi,Wu:1998wu}, and concluded that the $\Omega(2012)^-$ is not suitable
to be interpreted as a $(\bar{K}\Xi(1530))^-$ molecular state.

A theoretical study of the $\Omega(2012)^-$ resonance in the nonleptonic weak decays
$\Omega_{c}^{0} \to \pi^+ \bar{K} \Xi(1530)(\eta\Omega) \to \pi^{+} (\bar{K} \pi \Xi)^{-}$
and $\pi^+ (\bar{K} \Xi)^-$ via final-state interactions of the $\bar{K}\Xi(1530)$ and
$\eta\Omega$ pairs has been reported~\cite{Omega_c}. The authors found that the
$\Omega_{c}^{0} \to \pi^+ (\bar{K}\pi \Xi)^-$ decay is not well suited to study the
$\Omega(2012)^-$ because the dominant contribution is from the $\Omega_{c}^{0} \to \pi^+ (\bar{K} \Xi(1530))^-$
decay at tree level, and this will not contribute to the production of the $\Omega(2012)^-$.
On the other hand, they predicted that the $\Omega(2012)^-$ would be visible
in the $(\bar{K}\Xi)^-$ invariant mass spectrum of the $\Omega_{c}^{0} \to \pi^+ (\bar{K} \Xi)^-$
decay. It is clear that observing the $\Omega(2012)^-$ in different production mechanisms
can not only further confirm its existence but also yield important information that can
increase the understanding of its internal structure.

In this paper, we search for the $\Omega(2012)^-$ in the decay
$\Omega_{c}^{0} \to \pi^+\Omega(2012)^- \to \pi^+ (\bar{K}\Xi)^{-}$.
We first perform analyses separately for the two isospin modes
($\Omega_{c}^0 \to \pip \Omega(2012)^- \to \pi^+ K^- \Xi^0$/$\pi^+ K_S^0 \Xi^-$),
and then combine them for further analysis. Throughout this paper,
inclusion of charge-conjugate mode is implicitly assumed.

This analysis is based on data collected at or near the $\ones$, $\twos$,
$\threes$, $\fours$, and $\fives$ resonances by the Belle detector~\cite{detector1, detector2}
at the KEKB asymmetric-energy $e^+e^-$ collider~\cite{collider1, collider2}.
The total data sample corresponds to an integrated luminosity of 980~fb$^{-1}$~\cite{detector2}.
The Belle detector was a large-solid-angle magnetic spectrometer consisting of
a silicon vertex detector,
a 50-layer central drift chamber (CDC), an array of aerogel threshold Cherenkov counters (ACC),
a barrel-like arrangement of time-of-flight scintillation counters (TOF), and an electromagnetic
calorimeter comprising CsI(TI) crystals (ECL) located inside a superconducting solenoid coil that
provides a $1.5~\hbox{T}$ magnetic field. An iron flux return comprising resistive plate chambers
located outside the coil was instrumented to detect $K^{0}_{L}$ mesons and to identify muons.
A detailed description of the Belle detector can be found in Refs.~\cite{detector1, detector2}.

Monte Carlo (MC) simulated signal events are generated using {\sc EvtGen}~\cite{evtgen}
to optimize the signal selection criteria and calculate the reconstruction
efficiencies; $\EE \to c\bar{c}$ events are simulated using {\sc PYTHIA}~\cite{pythia},
where one of the two charm quarks hadronizes into an $\Omega_{c}^0$ baryon. Both
$\Omega_c^0 \to \pi^+ \Omega(2012)^-$ and $\Omega(2012)^- \to K^-\Xi^0/ K_S^0 \Xi^-$
decays are isotropic in the rest frame of the parent particle. We also generate the signal
MC events of $\Omega_c^0  \to  \pip K^-\Xi^0/ \pip K_S^0 \Xi^-$ decays with a phase-space
model to estimate the reconstruction efficiencies of the reference modes. The simulated
events are processed with a detector simulation based on {\sc GEANT3}~\cite{geant}.
Inclusive MC samples of $\Upsilon(1S,2S,3S)$ decays, $\fours \to B^{+}B^{-}/B^{0}\bar{B}^{0}$,
$\fives \to B_{s}^{(*)} \bar{B}_{s}^{(*)}$, and $\EE \to q\bar{q}$ $(q=u,~d,~s,~c)$
at center-of-mass (C.M.) energies of 10.520, 10.580, and 10.867~GeV corresponding to two
times the integrated luminosity of data are used to optimize the signal selection criteria and
to check possible peaking backgrounds~\cite{zhou}.

The impact parameters of the charged particle tracks, except for those
of the decay products of $K_S^0$, $\Lambda$, and $\Xi^-$, measured with respect to the
nominal interaction point (IP), are required to be less than 0.2~cm perpendicular to the
beam direction, and less than 1~cm parallel to it. For the particle identification (PID)
of a well-reconstructed charged track, information from different detector subsystems,
including specific ionization in the CDC, time measurement in the TOF, and the
response of the ACC, is combined to form a likelihood $\mathcal{L}_{i}$~\cite{pidcode}
for particle species $i$, where $i$ =  $K$, $\pi$, or $p$. Kaon candidates are defined
as those with $\mathcal{L}_{K}/(\mathcal{L}_K+\mathcal{L}_p) > 0.8$ and $\mathcal{L}_{K}/
(\mathcal{L}_K+\mathcal{L}_\pi) > 0.8$, which is approximately 87\% efficient.
For protons the requirements are $\mathcal{L}_{p}/(\mathcal{L}_{p}+\mathcal{L}_K) > 0.2$
and $\mathcal{L}_{p}/(\mathcal{L}_{p}+\mathcal{L}_\pi) > 0.2$, while for charged pions
$\mathcal{L}_{\pi}/(\mathcal{L}_{\pi}+\mathcal{L}_K) > 0.2$ and
$\mathcal{L}_{\pi}/(\mathcal{L}_{\pi}+\mathcal{L}_p) > 0.2$;
these requirements are approximately 99\% efficient.

An ECL cluster is taken as a photon candidate if it does not match the extrapolation
of any charged track. The $\pi^0$ candidates are reconstructed from two photons having
energy exceeding 30~MeV in the barrel or 50~MeV in the endcaps. The reconstructed
invariant mass of the $\pi^0$ candidate is required to be within 10.8~MeV/$c^2$ of the
$\pi^0$ nominal mass~\cite{PDG}, corresponding to approximately twice
the resolution ($\sigma$). To reduce the large combinatorial backgrounds, the momentum
of the $\pi^0$ candidate is required to exceed 200~MeV/$c$~\cite{Omega}. $\Lambda$
candidates are reconstructed from $p\pi^-$ pairs with a production vertex significantly
separated from the IP, and a reconstructed invariant mass within 3.5~MeV/$c^2$ of the
$\Lambda$ nominal mass~\cite{PDG} ($\sim$3$\sigma$).

The $\Xi^0 \to \Lambda \pi^0$ reconstruction is performed as follows.
The selected $\Lambda$ candidate is combined with a $\pi^0$ to form a $\Xi^0$ candidate,
and then taking the IP as the point of origin of the $\Xi^0$, the sum of the $\Lambda$
and $\pi^0$ momenta is taken as the momentum vector of the $\Xi^0$ candidate. The intersection of this
trajectory with the reconstructed $\Lambda$ trajectory is then found and this position
is taken as the decay location of the $\Xi^0$ baryon. The $\pi^0$ is then refit using this
location as its point of origin. Only those combinations with the decay location
of the $\Xi^0$ indicating a positive $\Xi^0$ path length of greater than 2~cm but
less than the distance between the $\Lambda$ decay vertex and the IP are retained~\cite{Omega}.
The $\Xi^-$ candidate is reconstructed by combining a $\Lambda$ candidate with a $\pim$.
The vertex formed from the $\Lambda$ and $\pim$ is required to be at least 0.35~cm
from the IP, to have a shorter distance from the IP than the $\Lambda$ decay vertex,
and to signify a positive $\Xi^-$ flight distance~\cite{Omega}.

The $K_{S}^{0}$ candidates are first reconstructed from pairs of oppositely charged tracks,
which are treated as pions, with a production vertex significantly separated from the average
IP, and then selected using an artificial neural network~\cite{neural} based on two sets of input
variables~\cite{input}.

The $\Xi^0$ and $\Xi^-$ are kinematically constrained to their nominal masses~\cite{PDG},
and then combined with a $K^-$ or $K_S^0$ to form an $\Omega(2012)^-$ candidate. Finally, the
reconstructed $\Omega(2012)^-$ candidate is combined with a $\pi^+$ to form an $\Omega_c^0$ candidate.
To improve the momentum resolution and suppress the backgrounds, a vertex fit (the IP is not included
in this vertex) is performed for the $\pi^+ (\bar{K} \Xi)^-$ final state, and then
$\chi^{2}_{\rm vertex} < 20$ is required, corresponding to an efficiency exceeding 90\%.

To reduce combinatorial backgrounds, especially from $B$-meson decays, the scaled momentum
$x_{p} = p^{*}_{\Omega_{c}^0}$/$p_{\rm max}$ is required to be larger than 0.6.
Here, $p^{*}_{\Omega_{c}^0}$ is the momentum of $\Omega_{c}^0$ candidates
in the $\EE$ C.M.\ frame, and $p_{\rm max}=\sqrt{E^2_{\rm beam}-M_{\Omega_c^0}^2 c^4}/c$,
where $E_{\rm beam}$ is the beam energy in the $\EE$ C.M.\ frame
and $M_{\Omega_{c}^0}$ is the invariant mass of $\Omega_{c}^0$ candidates. This criterion
is optimized by maximizing the Punzi figure of merit = $S/(3/2 + \sqrt{B}$)~\cite{Punzi},
where  $S$ is the number of expected $\Omega_c^0 \to \pi^+ \Omega(2012)^- \to \pi^+ (\bar{K} \Xi)^-$
signal events from signal MC samples, by performing a two-dimensional (2D) maximum-likelihood fit to $M((\bar{K}\Xi)^-)$
and $M(\pi^+ \Omega(2012)^-)$ distributions and assuming $\sigma(\EE \to \Omega_{c}^0 + anything) \times
\BR(\Omega_c^0 \to \pi^+ \Omega(2012)^-) \times \BR(\Omega(2012)^- \to (\bar{K} \Xi)^-)$ = 10~fb, and $B$ is
the number of background events from a 2D fit from inclusive MC samples.

Reconstructed invariant masses for $\Xi^0$, $K_S
^0$, and $\Xi^-$ candidates are required to be within
7.0, 7.0, and 3.5~MeV/$c^2$ of the corresponding nominal masses~\cite{PDG} ($>94\%$
signal events are retained for each intermediate state), respectively. These requirements
are optimized using the same method as was used for scaled momentum.

Finally, if there are multiple $\Omega_{c}^0$ candidates in an event,
all the combinations are retained for further analysis. The fractions of events with multiple combinations
for $\Omega_{c}^0 \to \pip \Omega(2012)^- \to \pip K^- \Xi^0$ and $\Omega_c^0 \to
\pip \Omega(2012)^- \to \pip K_S^0 \Xi^-$ decays are 2.4\% and 0.8\%, respectively,
which are consistent with the signal MC expectations.

\begin{figure}[htbp]
	\begin{center}
		\includegraphics[width=4.5cm]{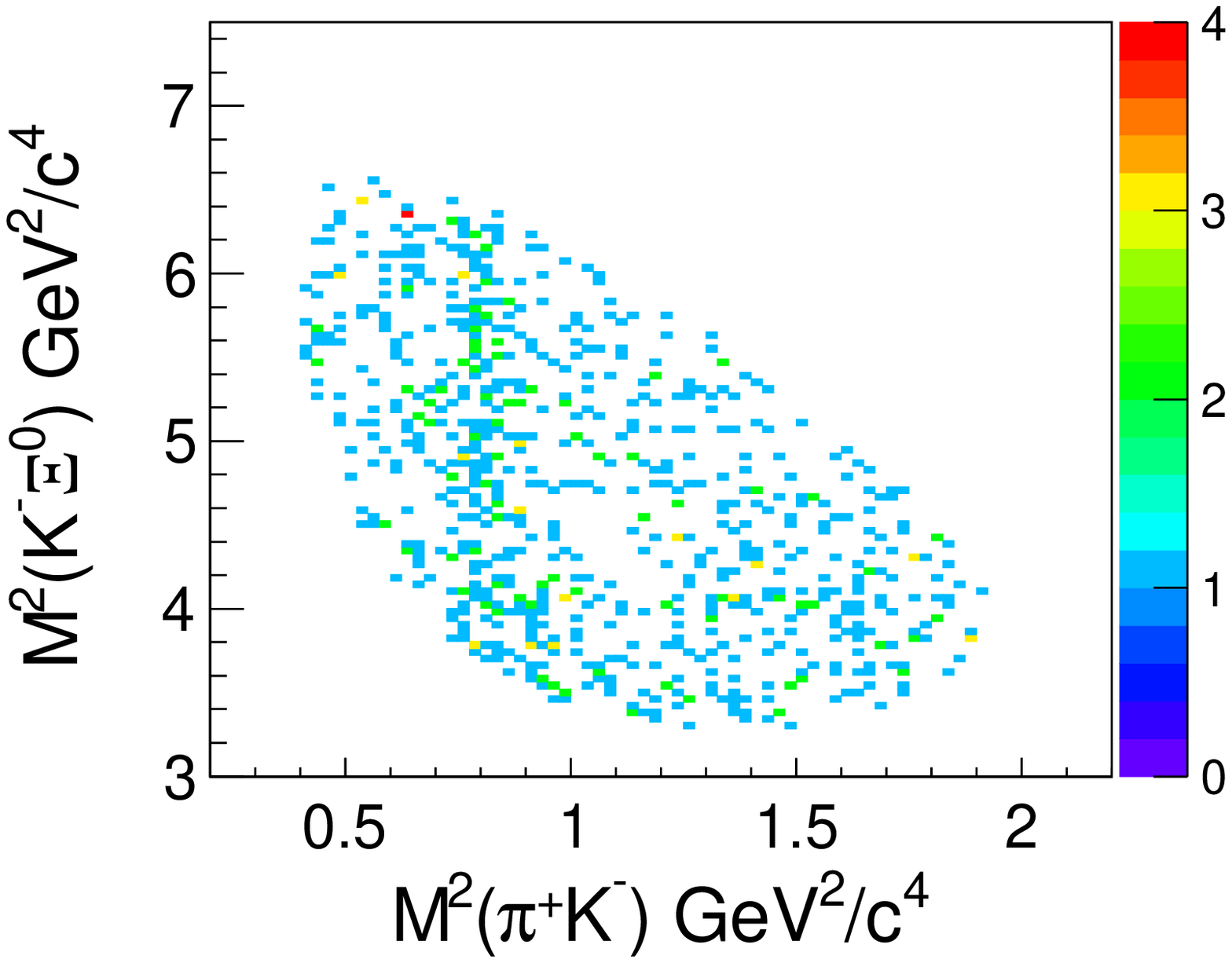}\hspace{-0.55cm}
		\includegraphics[width=4.5cm]{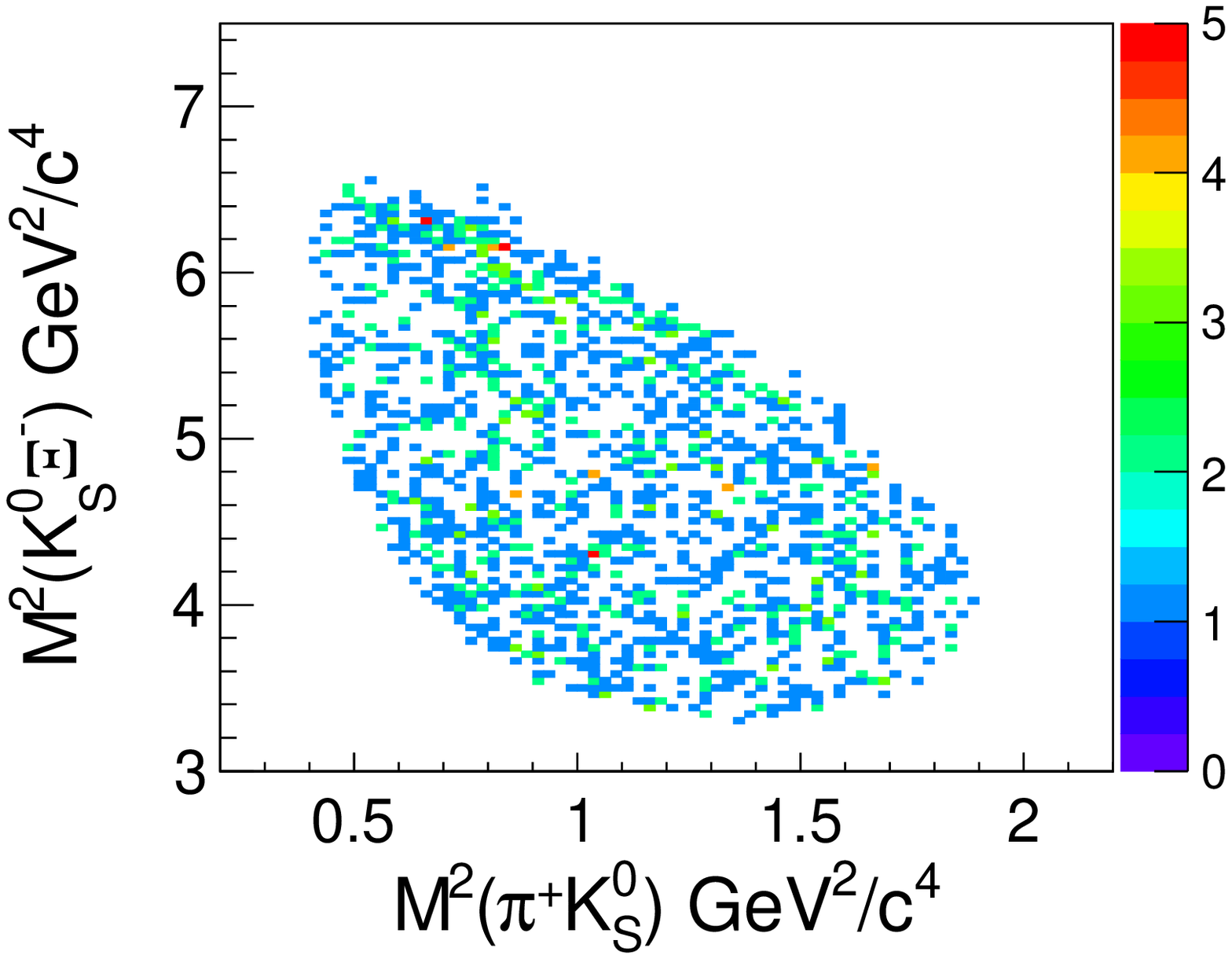}
		\put(-240,78){\bf (a)} \put(-125,78){\bf (b)}
		\caption{The Dalitz plots of (a) $M^{2}(K^-\Xi^0)$ versus $M^{2}(\pip K^-)$ and (b)
		$M^{2}(K_S^0 \Xi^-)$ versus $M^{2}(\pip K_S^0)$ from selected $\Omega_{c}^0 \to \pip K^- \Xi^0$
		and $\Omega_c^0 \to \pip K_S^0 \Xi^-$ candidates.}\label{fig1}
	\end{center}
\end{figure}

After applying the aforementioned event selection criteria, the Dalitz
plots of $M^{2}(K^-\Xi^0)$ versus $M^{2}(\pip K^-)$ and $M^{2}(K_S^0\Xi^-)$ versus
$M^{2}(\pip K_S^0)$ in the $\Omega_c^0$ signal region are shown in Fig.~\ref{fig1}, where the reconstructed invariant mass of $\Omega_c^0$
candidates is required to be within 15~MeV/$c^2$ of the $\Omega_c^0$ nominal mass~\cite{PDG} ($\sim$2.5$\sigma$).

\begin{figure}[htbp]
	\begin{center}
		\includegraphics[width=4.5cm]{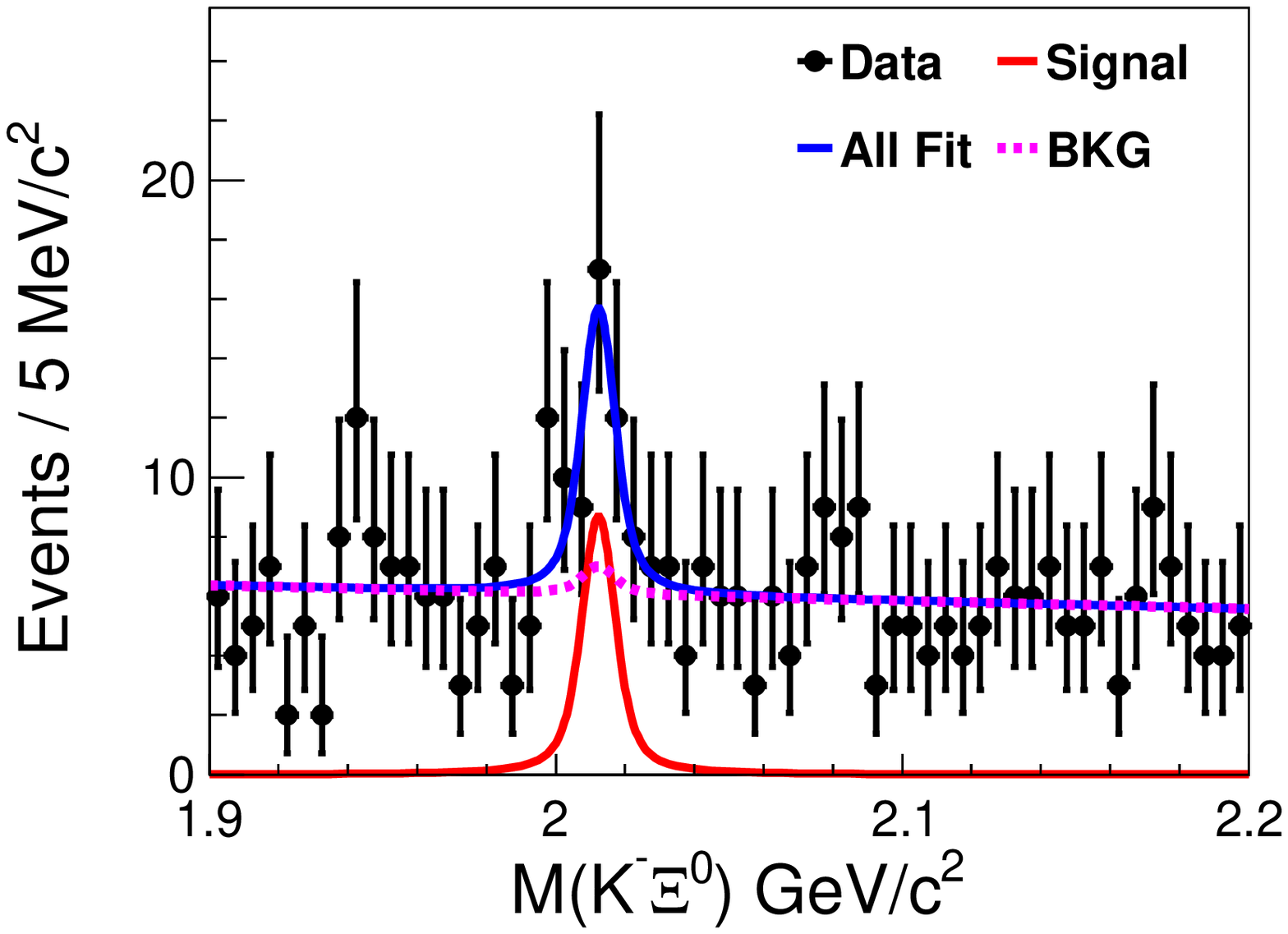}\hspace{-0.55cm}
		\includegraphics[width=4.5cm]{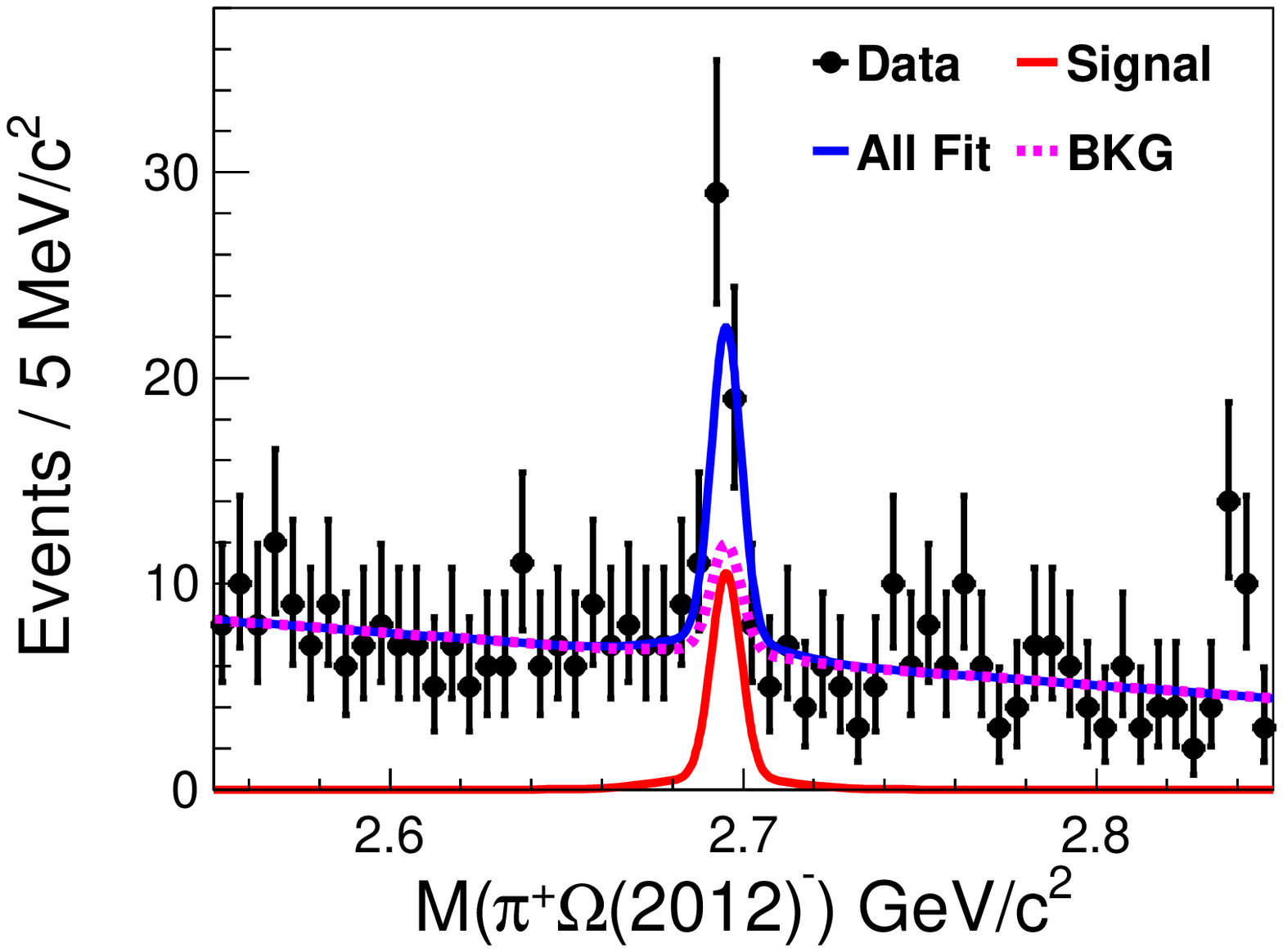}
		\put(-210,70){\bf (a1)} \put(-95,70){\bf (b1)}
		
		\includegraphics[width=4.5cm]{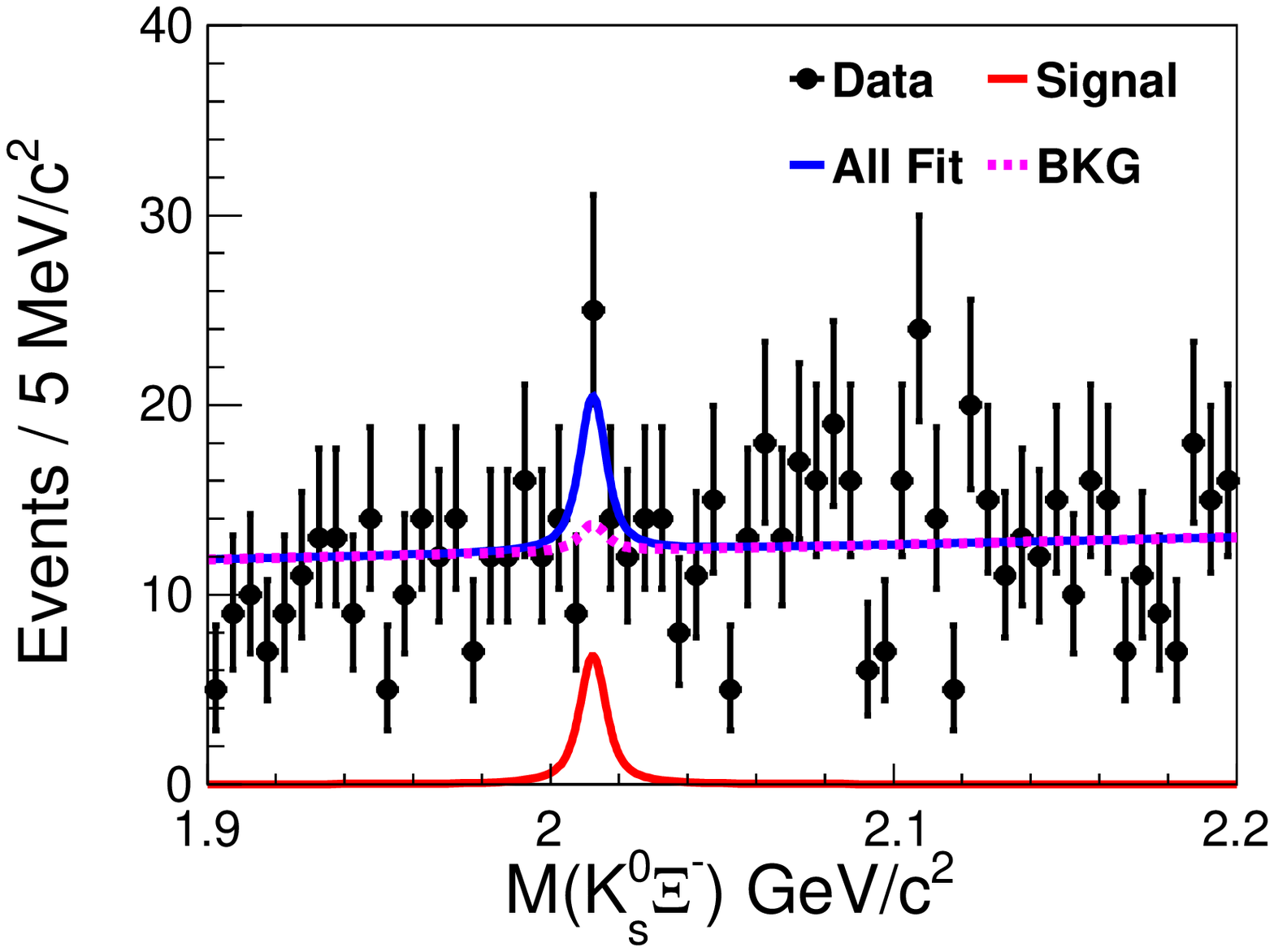}\hspace{-0.55cm}
		\includegraphics[width=4.5cm]{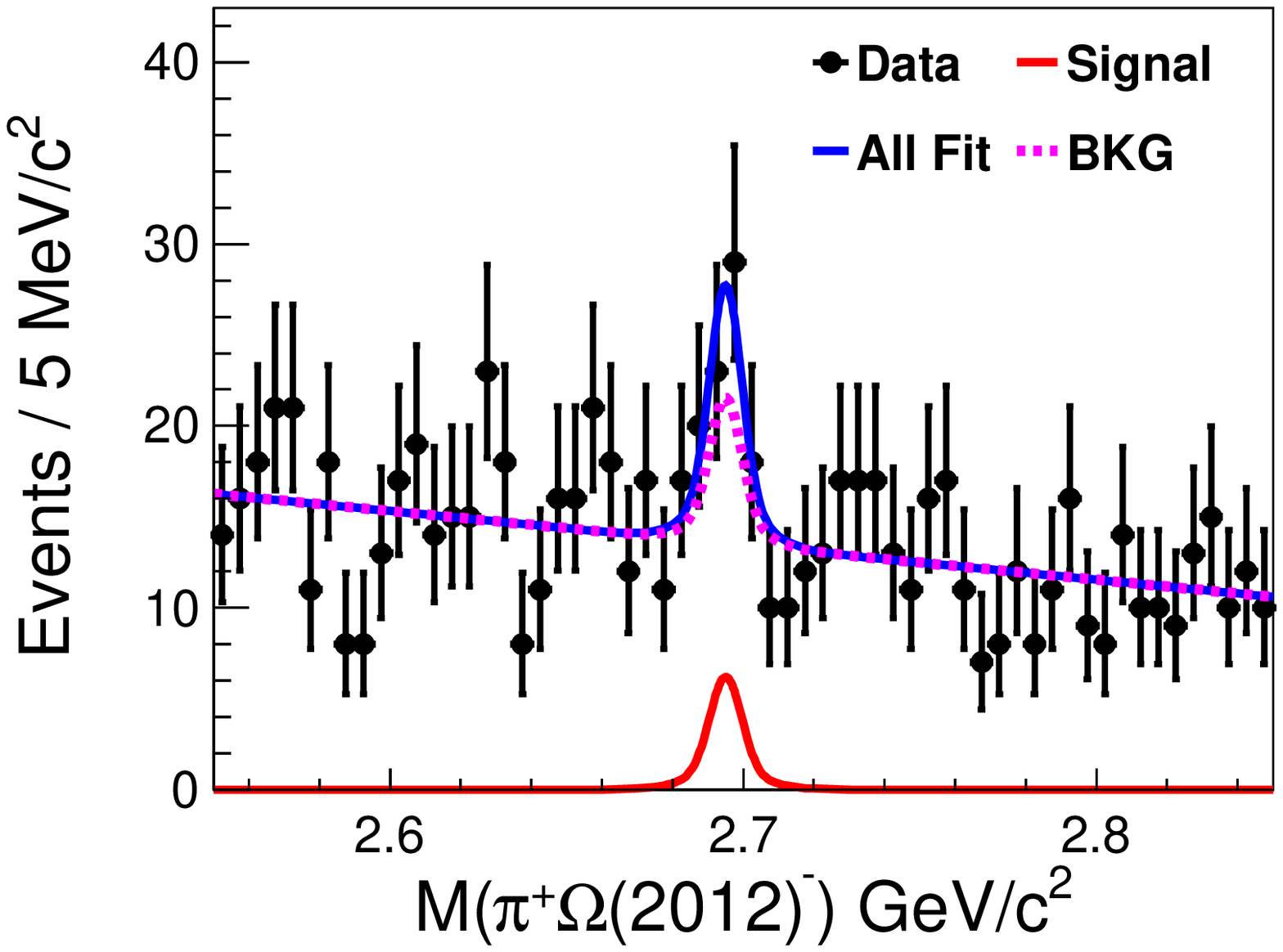}
		\put(-210,70){\bf (a2)} \put(-95,70){\bf (b2)} 	
		\caption{The 1D projections of the 2D fits of (a) $M(K^- \Xi^0)$/$M(K_S^0 \Xi^-)$ and (b) $M(\pi^+ \Omega(2012)^-)$ distributions for (1) $\Omega_{c}^0 \to \pi^+ \Omega(2012)^-\to \pip K^-\Xi^0$ and (2) $\Omega_{c}^0 \to \pi^+ \Omega(2012)^-\to \pip K_S^0 \Xi^-$ decays in data. All components are indicated in the legends and described in the text.}\label{fig2}
	\end{center}
\end{figure}

To extract the $\Omega(2012)^-$ signal events from $\Omega_{c}^0$ decay,
we perform a 2D unbinned maximum-likelihood fit to $M(K^-\Xi^0$)/$M(K_S^0 \Xi^-)$
and $M(\pi^+ \Omega(2012)^-)$ distributions. The 2D fitting function $f(M_1,M_2)$
is expressed as
\begin{align*}
    f(M_1,M_2)= N^{\rm sig}_{\rm ss} s_1(M_1) s_2(M_2) &+
	N^{\rm bg}_{\rm sb} s_1(M_1) b_2(M_2) \\
	+ N^{\rm bg}_{\rm bs} b_1(M_1)
	s_2(M_2)  &+ N^{\rm bg}_{\rm bb}  b_1(M_1)
	b_2(M_2),
\end{align*}
where $s_1 (M_1)$ and $b_1(M_1)$ are the signal and background probability density functions (PDFs)
for the $M(K^- \Xi^0)$/$M(K_S^0 \Xi^-)$ distributions, respectively, and $s_2 (M_2)$ and $b_2(M_2)$ are the
corresponding PDFs for the $M(\pi^+ \Omega(2012)^-)$ distributions. Here, $N^{\rm sig}_{\rm ss}$ is the number of
signal events, $N^{\rm bg}_{\rm sb}$ and $N^{\rm bg}_{\rm bs}$
denote the numbers of peaking background events in $M(K^- \Xi^0)$/$M(K_S^0 \Xi^-)$ and $M(\pi^+ \Omega(2012)^-)$
distributions, respectively, and $N^{\rm bg}_{\rm bb}$ is the number of combinatorial background events
both for $\Omega(2012)^-$ and $\Omega_{c}^0$ candidates. The signal shapes ($s_1 (M_1)$ and $s_2 (M_2)$) of
$\Omega(2012)^-$ and $\Omega_{c}^0$ candidates are described by a Breit-Wigner (BW) function convolved with
a Gaussian function and a double-Gaussian function, respectively, and first-order polynomial functions
represent the backgrounds ($b_1 (M_1)$ and $b_2 (M_2)$). The values of signal PDF parameters are fixed
to those obtained from the fits to the corresponding simulated signal distributions. The values of the
background shape parameters are allowed to float in the fit. The one-dimensional (1D) projections of
$M(K^-\Xi^0)/M(K_S^0\Xi^-)$ in the $\Omega_{c}^0$ signal region and $M(\pip \Omega(2012)^-)$ in the $\Omega(2012)^-$ signal region from 2D fits are shown in Fig~\ref{fig2}.  The signal regions of $\Omega(2012)^-$ and $\Omega_c^0$
candidates are defined as $|M(K^- \Xi^0)$/$M(K_S^0 \Xi^-)-m(\Omega(2012)^-)|$ $<$ 20~MeV/$c^2$ ($\sim$2.5$\sigma$) and
$|M(\pi^+ \Omega(2012)^-)-m(\Omega_{c}^0)|$ $<$ 15~MeV/$c^2$ ($\sim$2.5$\sigma$),
respectively, where $m(\Omega(2012)^-)$ and $m(\Omega_{c}^0)$ are the nominal
masses of $\Omega(2012)^-$ and $\Omega_c^0$~\cite{PDG}. The numbers of fitted $\Omega_{c}^0 \to \pi^+
\Omega(2012)^-\to \pip K^-\Xi^0$ and $\Omega_{c}^0 \to \pi^+ \Omega(2012)^-\to \pip K_S^0 \Xi^-$ signal
events are $28.3 \pm 8.9$ and $17.9 \pm 8.9$ with statistical significances of $4.0\sigma$ and $2.3\sigma$,
respectively. Here, the statistical significances are defined as
$\sqrt{-2\ln(\mathcal{L}_{0}/\mathcal{L}_\text{max})}$, where $\mathcal{L}_{0}$ and
$\mathcal{L}_\text{max}$ are the maximized likelihoods without and with a signal component, respectively.

\begin{figure}[htbp]
	\begin{center}
		\includegraphics[width=4.5cm]{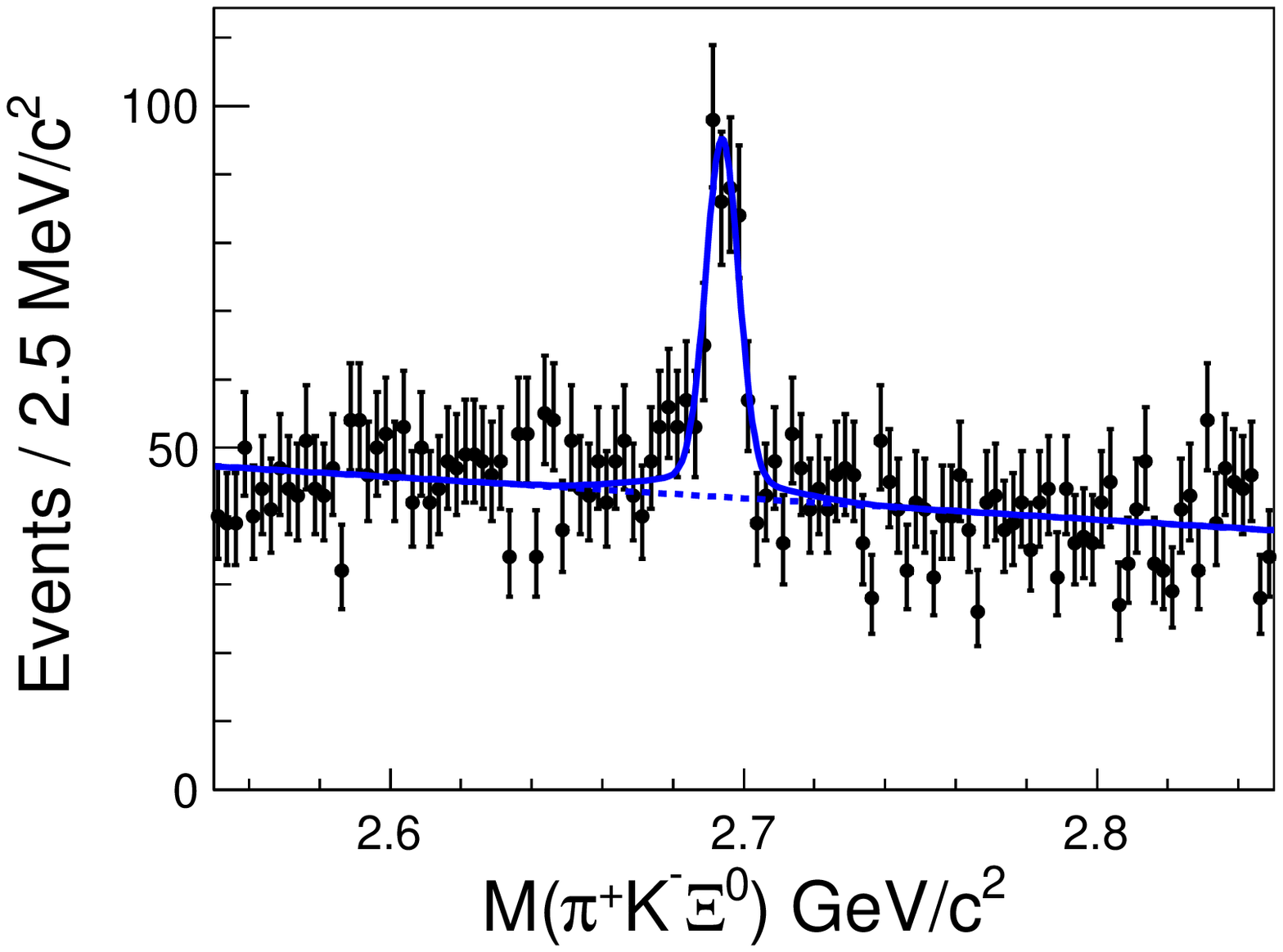}\hspace{-0.55cm}
		\includegraphics[width=4.5cm]{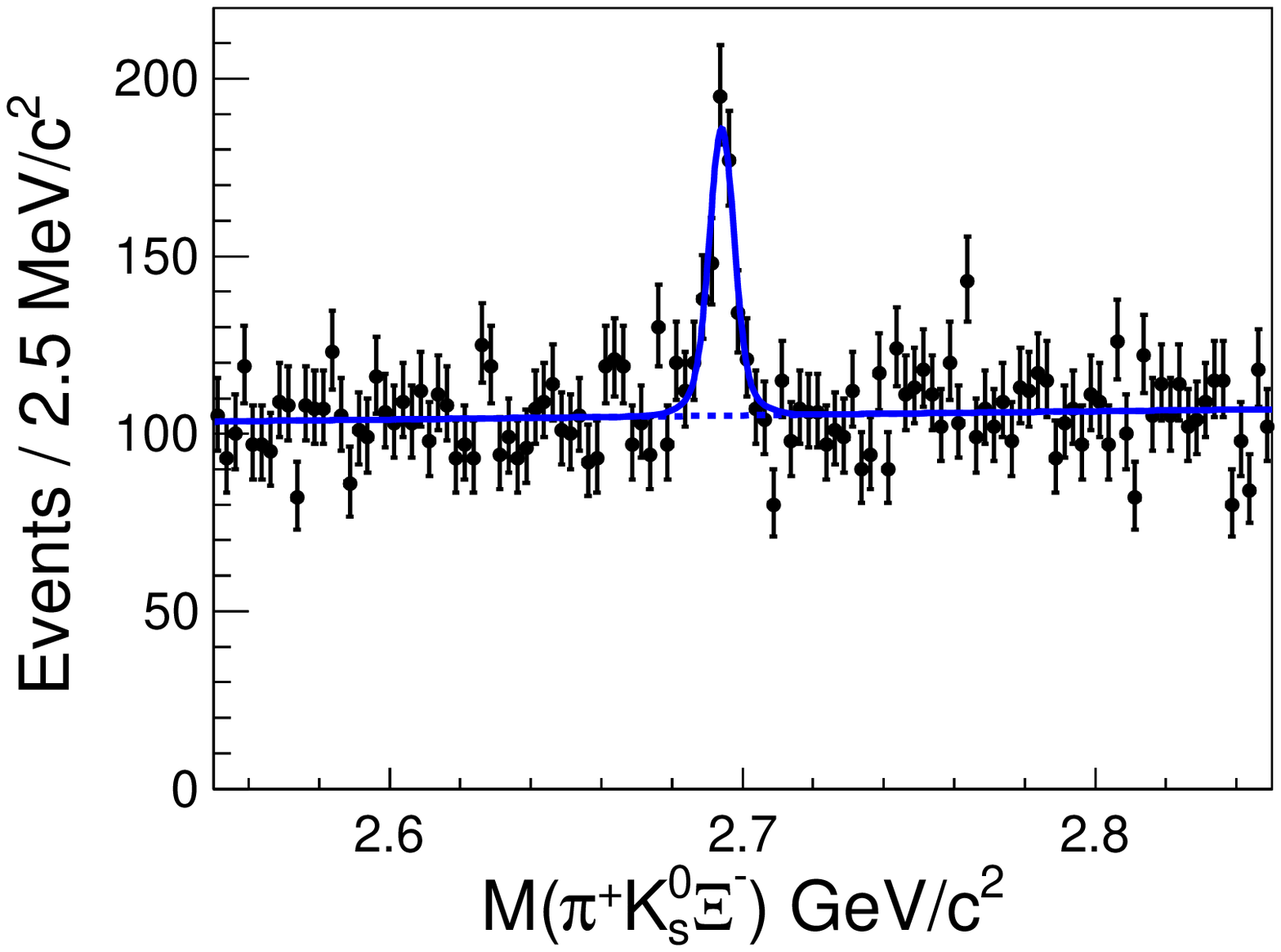}
        \put(-210,70){\bf (a)} \put(-95,70){\bf (b)} 	
		\caption{The (a) $M(\pi^+ K^- \Xi^0)$ and (b) $M(\pi^+ K_S^0 \Xi^-)$ distributions in data.
			The blue solid curves show the best-fit results, and the blue dashed
			curves show the fitted backgrounds.}\label{fig3}
	\end{center}
\end{figure}

For $\Omega_{c}^{0} \to \pi^+ K^- \Xi^0$ and $\Omega_{c}^{0} \to \pi^+ K_S^0 \Xi^-$
decays, the $M(\pi^+ K^- \Xi^0)$ and $M(\pi^+ K_S^0 \Xi^-)$ distributions are shown
in Fig.~\ref{fig3}, together with the fitted results. The signal shapes of $\Omega_c^0$ are described
by double-Gaussian functions, where the parameters are fixed to those obtained from the fits to the
corresponding simulated signal distributions. The backgrounds are parametrized by first-order polynomial
functions. The fitted $\Omega_c^0 \to \pi^+ K^- \Xi^0$ and $\Omega_c^0 \to \pi^+ K_S^0 \Xi^-$ signal yields
are $279 \pm 27$ and $317 \pm 32$, respectively.

The branching fraction ratios are calculated according to the formulae
\begin{align*}
\cal{R}_{\rm 1} = & \frac{\BR(\Omega_{c}^0 \to \pi^+ \Omega(2012)^-)\BR(\Omega(2012)^- \to K^- \Xi^0)}{\BR(\Omega_{c}^0 \to \pi^+ K^- \Xi^0)} \\ = & \frac{N_{\pi^+\Omega(2012)^-(\to K^- \Xi^0)}^{\rm obs} \times \epsilon_{\pi^+ K^- \Xi^0}}{N_{\pi^+ K^- \Xi^0}^{\rm obs} \times \epsilon_{\pi^+\Omega(2012)^-(\to K^- \Xi^0)}} \\ = & (9.6 \pm 3.2(\rm stat.) \pm 1.8(\rm syst.))\%,
\end{align*}
and
\begin{align*}
\cal{R}_{\rm 2} = &\frac{\BR(\Omega_{c}^0 \to \pi^+ \Omega(2012)^-)\BR(\Omega(2012)^- \to \bar{K}^0 \Xi^-)}
	{\BR(\Omega_{c}^0 \to \pi^+ \bar{K}^0 \Xi^-)} \\
	= &\frac{N_{\pi^+\Omega(2012)^-(\to  K_S^0 \Xi^-)}^{\rm obs} \times \epsilon_{\pi^+  K_S^0 \Xi^-}}{N_{\pi^+  K_S^0 \Xi^-}^{\rm obs} \times \epsilon_{\pi^+\Omega(2012)^-(\to  K_S^0 \Xi^-)}} \\ = & (5.5 \pm 2.8(\rm stat.) \pm 0.7(\rm syst.))\%.
\end{align*}
Here, $N_{\pi^+\Omega(2012)^-(\to K^- \Xi^0)}^{\rm obs}$, $N_{\pi^+\Omega(2012)^-(\to  K_S^0 \Xi^-)}^{\rm obs}$, $N_{\pi^+ K^- \Xi^0}^{\rm obs}$, and $N_{\pi^+  K_S^0 \Xi^-}^{\rm obs}$ are the fitted signal yields in the decay modes $\Omega_{c}^0 \to \pi^+ \Omega(2012)^- \to \pip K^- \Xi^0$,  $\Omega_{c}^0 \to \pi^+ \Omega(2012)^- \to \pip K_S^0 \Xi^-$, $\Omega_{c}^0 \to \pip K^- \Xi^0$, and $\Omega_{c}^0 \to \pi^+  K_S^0 \Xi^-$, respectively; $\epsilon_{\pi^+\Omega(2012)^-(\to K^- \Xi^0)}$, $\epsilon_{\pi^+\Omega(2012)^-(\to  K_S^0 \Xi^-)}$, $\epsilon_{\pi^+ K^- \Xi^0}$, and $\epsilon_{\pi^+  K_S^0 \Xi^-}$
are the corresponding reconstruction efficiencies, which are obtained from the signal MC simulations and are listed in Table~\ref{tab:sigeff1}. The systematic uncertainties are discussed below.

\begin{table}[htbp]
	\caption{\label{tab:sigeff1} Summary of the fitted signal yields ($N^{\rm obs}$) and reconstruction efficiencies ($\epsilon$).
	All the uncertainties here are statistical only.}
	\begin{tabular}{ccc}
		\hline
		\hline		
		Mode & $N^{\rm obs}$ & $\epsilon$(\%)  \\		
		\hline
		$\Omega_{c}^0 \to \pi^+ \Omega(2012)^- \to \pip K^-  \Xi^0$     &  28.3 $\pm$ 8.9   & 3.59 \\
		$\Omega_{c}^0 \to \pi^+ \Omega(2012)^- \to \pip K_S^0 \Xi^-$    &  17.9 $\pm$ 8.9   & 7.68  \\
		$\Omega_{c}^0 \to \pip K^- \Xi^0$                               &   279 $\pm$ 27    & 3.41  \\
		$\Omega_{c}^0 \to \pi^+  K_S^0 \Xi^-$                           &   317 $\pm$ 32    & 7.41  \\
		\hline
		\hline
	\end{tabular}
\end{table}

From these fitted signal yields and reconstruction efficiencies,
and the intermediate state branching fractions of $\Omega_c^0 \to \pip \Omega(2012)^- \to \pip K^- \Xi^0$
and $\Omega_c^0 \to \pip \Omega(2012)^- \to \pip K_S^0 \Xi^-$ decays~\cite{PDG}, the branching fraction ratio
$\BR(\Omega(2012)^- \to K^- \Xi^0)$/$\BR(\Omega(2012)^- \to \bar{K}^0 \Xi^-)$ is determined to be 1.19 $\pm$ 0.70(stat.),
which is consistent with the expectation of isospin symmetry and the previously measured value of $1.2 \pm 0.3$ by Belle~\cite{Omega}.

\begin{figure}[htbp]
	\begin{center}
		\includegraphics[width=4.5cm]{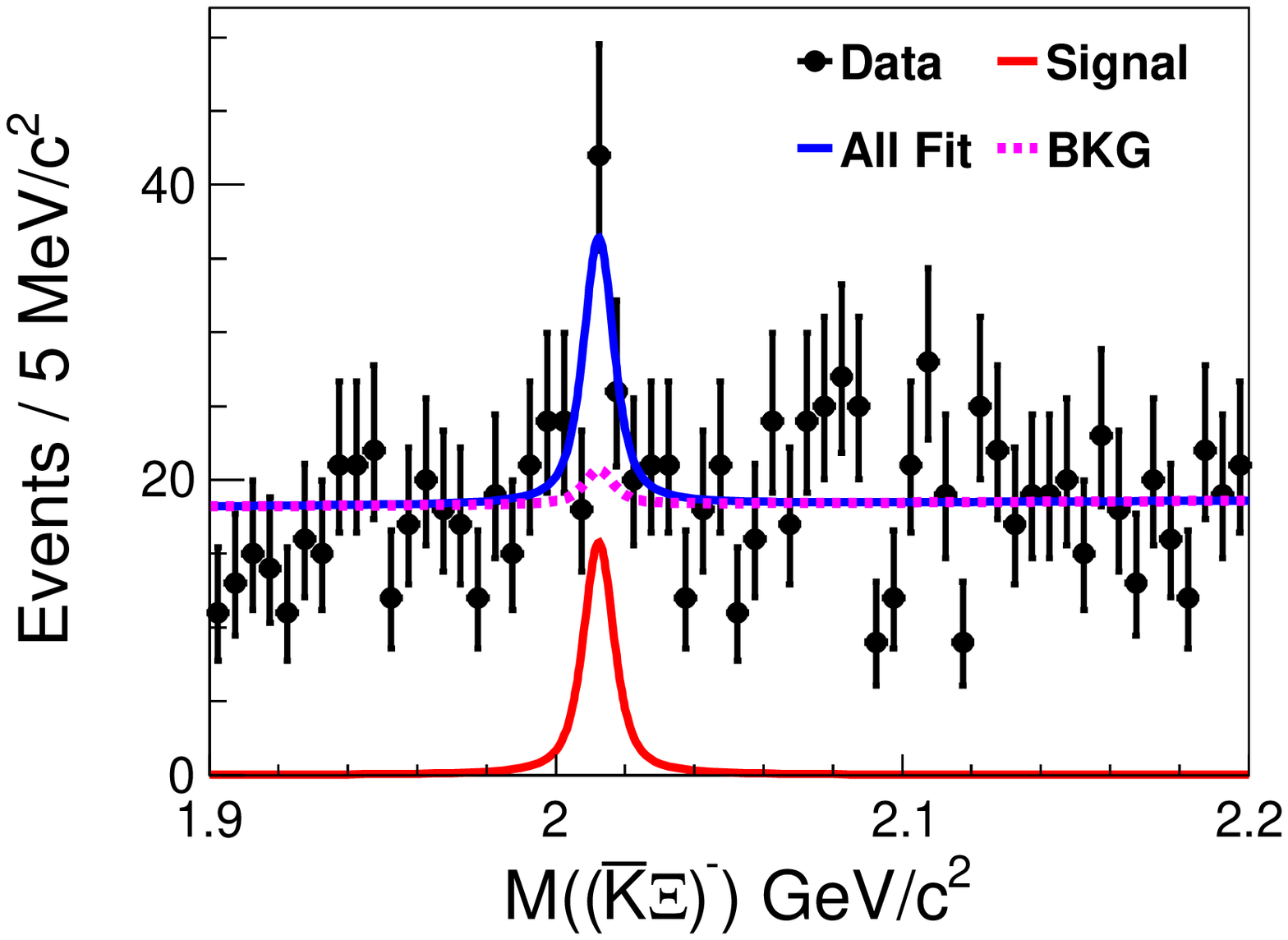}\hspace{-0.55cm}
		\includegraphics[width=4.5cm]{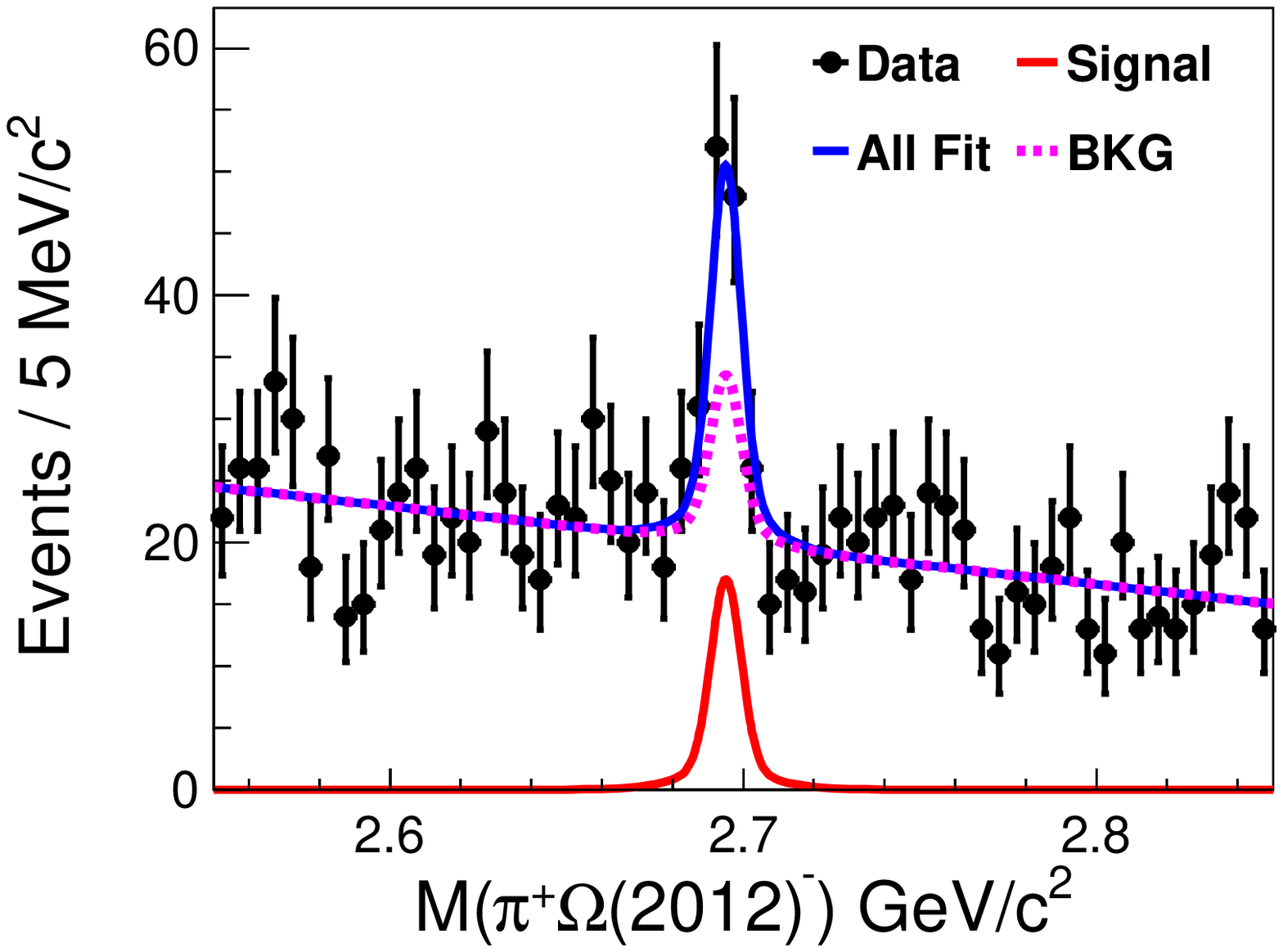}
		\put(-210,70){\bf (a)} \put(-95,70){\bf (b)}
		\caption{The 1D projections of the 2D simultaneous fit of (a) $M((\bar{K} \Xi)^-)$ and (b)
		 $M(\pi^+ \Omega(2012)^-)$ distributions in data. All components are
		 indicated in the legends and described in the text.}\label{fig4}
	\end{center}
\end{figure}

Assuming $\BR(\Omega(2012)^- \to K^- \Xi^0)$ = $\BR(\Omega(2012)^- \to \bar{K}^0 \Xi^-)$
based on isospin symmetry, the ratio of the expected signal yields of $\Omega_{c}^0
\to \pi^+ \Omega(2012)^- \to \pi^+ K^- \Xi^0$ and $\Omega_{c}^0 \to \pi^+ \Omega(2012)^- \to
\pi^+ K_S^0 \Xi^-$ decays is 57.1\%:42.9\% after considering the products of detection efficiency and intermediate-state branching fractions $\epsilon_{i} \BR_{i}$ ($i=1,~2$), where $\epsilon_{1}$ and $\epsilon_{2}$  are the corresponding detection efficiencies, $\BR_{1} = \BR(\Xi^0 \to \Lambda \pi^0) \times \BR(\pi^0 \to \gamma\gamma)$, and $\BR_{2} = \BR(\Xi^- \to \Lambda \pim) \times \BR(\bar{K}^0 \to K_{S}^0) \times \BR(K_{S}^{0} \to \pip\pim)$~\cite{PDG}.
We perform a 2D unbinned maximum-likelihood simultaneous fit to $M((\bar{K} \Xi)^-)$ and $M(\pi^+ \Omega(2012)^-)$ distributions, where the ratio of the expected signal yields of two isospin modes is fixed to 57.1\%:42.9\%, and the functions used to describe the signal and background shapes are parameterized as before. The 1D projections of $M((\bar{K}\Xi)^-)$ in the $\Omega_{c}^0$ signal region and $M(\pip \Omega(2012)^-)$ in the $\Omega(2012)^-$ signal region
from the 2D simultaneous fit are shown in Fig.~\ref{fig4}, corresponding to a total signal yield of $46.6 \pm 12.3$. The statistical significance of the $\Omega(2012)^-$ signal in $\Omega_{c}^0 \to \pi^+ \Omega(2012)^- \to \pip (\bar{K}\Xi)^-$ decay is $4.6\sigma$. The fitting ranges and background shapes are the dominant systematic uncertainties
for the estimate of the signal significance. If the background shapes are replaced by second-order
polynomial functions and fitting ranges are changed, the $\Omega(2012)^-$ signal significance in
the simultaneous fit is reduced to 4.2$\sigma$ corresponding to a total signal yield of 44.7$\pm$12.4.
We take this value as the signal significance with systematic uncertainties included.

The $\Omega(2012)^-$ was first observed in data taken at the $\Upsilon(1S)$, $\Upsilon(2S)$,
and $\Upsilon(3S)$ resonances~\cite{Omega}. In order to make a statistically independent check of its existence,
we exclude these data sets from our sample and repeat the fitting procedure used to produce Fig.~\ref{fig4}.
The total number of signal events of $\Omega_{c}^0 \to \pi^+ \Omega(2012)^- \to \pip (\bar{K}\Xi)^-$ is $38.9 \pm 11.2$ in this reduced data sample which corresponds to an integrated luminosity of 949.5~fb$^{-1}$, and the statistical significance of the signal is 4.2$\sigma$. We prefer to use the entire data set for our investigation of the branching fractions of the $\Omega_c^0$.

The ratio of the branching fraction of $\Omega_{c}^0 \to \pi^+ \Omega(2012)^- \to \pi^+(\bar{K} \Xi)^{-}$
relative to that of $\Omega_{c}^0 \to \pi^+ \Omega^-$ decay is also calculated from the following formula
\begin{align*}
\cal{R}_{\rm 3} = &\frac{\BR(\Omega_{c}^0 \to \pi^+ \Omega(2012)^-) \times \BR(\Omega(2012)^- \to (\bar{K} \Xi)^{-})}
{\BR(\Omega_{c}^0 \to \pi^+ \Omega^-)} \notag \\= & \frac{N_{\rm sig.}^{\rm obs} \times \epsilon_{\pi^+ \Omega^-}}{N_{\pip \Omega^-}^{\rm obs} \times (f_1 \times \epsilon_1 \times \BR_{1} + f_2 \times \epsilon_{2} \times \BR_{2})} \notag \\ = & 0.220 \pm 0.059(\rm stat.) \pm 0.035(\rm syst.),
\end{align*}
where $N_{\rm sig.}^{\rm obs}$ is the fitted signal yield from the simultaneous fit in the decay
$\Omega_{c}^0 \to \pi^+ \Omega(2012)^- \to \pip (\bar{K}\Xi)^-$; $\epsilon_{1}$ and $\epsilon_{2}$ are the
corresponding reconstruction efficiencies from the signal MC simulations; according to isospin symmetry,
$f_{1} = \BR(\Omega(2012)^- \to K^- \Xi^0)/\BR(\Omega(2012)^- \to (\bar{K} \Xi)^-)$ = 0.5, $f_{2} = \BR(\Omega(2012)^- \to \bar{K}^0 \Xi^-)/\BR(\Omega(2012)^- \to (\bar{K} \Xi)^-)$ = 0.5; $\BR_{1}$ and $\BR_{2}$ are the corresponding products of secondary branching fractions defined above;
$N_{\pip \Omega^-}^{\rm obs}$  = 691 $\pm$ 29 and $\epsilon_{\pi^+ \Omega^-}$ = 10.08\%  are the number of signal events and detection efficiency of $\Omega_c^0 \to \pip \Omega^-$ decay taken from Ref.~\cite{Omega_c_belle}.

There are several sources of systematic uncertainties for the measurements
of branching fraction ratios $\cal{R}$$_{1}$, $\cal{R}$$_{2}$, and $\cal{R}$$_{3}$
as listed in Table~\ref{tab:errsum}, including detection-efficiency-related uncertainties, the
statistical uncertainty of the MC efficiency, the modeling of MC event generation,
the branching fractions of intermediate states,
the $\Omega(2012)^-$ resonance parameters,
the uncertainty in the $\Xi^0$ mass (as evaluated from the difference between the reconstructed
value and the world average value), as well as the overall fit uncertainty.

The detection-efficiency-related uncertainties include those for tracking efficiency (0.35\%
per track), PID efficiency (1.2\% per kaon, 1.0\% or 1.2\% per pion depending on the specific
decay mode), $K_S^0$ selection efficiency (1.7\%), as well as $\pi^0$ reconstruction efficiency
(2.25\%). For the measurements of $\cal{R}$$_{1}$ and $\cal{R}$$_{2}$, the detection-efficiency-related
sources can cancel. For the measurement of $\cal{R}$$_{3}$, the common sources of systematic
uncertainties such as $\Lambda$ selection cancel; to determine the total
detection-efficiency-related uncertainties, the above individual uncertainties from
different reconstructed modes ($\sigma_{i/\pi^+\Omega^-}$) are
added using the following standard error propagation formula
\begin{center}	
	$\sigma_{\rm DER}^{\cal{R}_{\rm 3}} = \sqrt{\frac{\Sigma_{i}(\W_{i}\times \sigma_{i})^2}{(\Sigma_{i}\W_{i})^2} + \sigma_{\pi^+\Omega^-}^{2}}$,
\end{center}
where $\W_{i}$ ($\W_{1}$ = $f_1\times\epsilon_{1}\times\BR_{1}$,
$\W_{2}$ = $f_2\times\epsilon_{2}\times\BR_{2}$) is the weight factor for the $i$-th ($i = 1,~2$)
mode of $\Omega_{c}^0 \to \pip\Omega(2012)^- \to \pip(\bar{K} \Xi)^-$ decays.
Assuming these sources are independent and adding them in quadrature, the final uncertainty
related to the reconstruction efficiency in the measurement of $\cal{R}$$_{3}$ is
2.2\%.

The MC statistical uncertainties are all 1.0\% or less. We assume that both
$\Omega_c^0 \to \pi^+ \Omega(2012)^-$ and $\Omega(2012)^- \to K^-\Xi^0/ K_S^0 \Xi^-$ decays
are isotropic in the rest frame of the parent particle, and a phase space model is used
to generate signal events. Since the signal efficiency is independent of the decay angular
distributions of $\pi^+$ in $\Omega_c^0$ C.M. and $K^-/K_S^0$ in $\Omega(2012)^-$ C.M.,
the model-dependent uncertainty has negligible effect on efficiency.
For the measurement of $\cal{R}$$_{3}$, the uncertainties from the
$\BR(\Xi^0 \to \Lambda \pi^0)$, $\BR(\Xi^- \to \Lambda \pi^-)$, $\BR(K_S^0 \to \pip \pim)$,
and $\BR(\pi^0 \to \gamma \gamma)$ are 0.012\%, 0.035\%, 0.072\%, and 0.035\%~\cite{PDG},
respectively, which are small and neglected. The uncertainties related to the mass and
width of $\Omega(2012)^-$ resonance are considered as diffierent sources, and are
estimated by changing the values of resonance mass and width by
$\pm$1$\sigma$ and refitting~\cite{Omega}.
The largest differences compared to the nominal fit results are added in quadrature as
systematic uncertainty. The uncertainty in the $\Xi^0$ mass is estimated by comparing the
signal yields of $\Omega_{c}^0 \to \pip \Omega(2012)^- \to \pip K^- \Xi^0/\pip (\bar{K} \Xi)^-$
for the case where the reconstructed $\Xi^0$ mass is
fixed at the found peak value versus the case where the mass is fixed at the nominal
mass~\cite{PDG}.

The systematic uncertainties associated with the fit range, background shape,
and mass resolution are considered as follows. To consider the uncertainty due to mass
resolution, we enlarge the mass resolution of signal by 10\% and take the difference in
the number of signal events as the systematic uncertainty. The order of the background polynomial
is replaced by a higher-order Chebyshev function and the fit range is changed. The largest
deviation compared to the nominal fit results is taken as the systematic uncertainty.
For each mode,
all the above uncertainties are summed in quadrature to obtain the total systematic uncertainty due to the fit.
Finally, the fit uncertainties of signal and reference modes are added in quadrature as total fit uncertainties in
the measurements of branching fraction ratios.

We estimate the uncertainty in $\cal{R}_{\rm 3}$ associated with the
ratio of the expected signal yields of the $\Omega_{c}^0 \to \pi^+ \Omega(2012)^- \to
\pi^+ K^- \Xi^0$ and $\Omega_{c}^0 \to \pi^+ \Omega(2012)^- \to \pi^+ K_S^0 \Xi^-$ decays
by constraining the ratio of $\BR( \Omega(2012)^- \to K^- \Xi^0)$:$\BR(\Omega(2012)^-
\to \bar{K}^0 \Xi^-)$ to 1.2:1~\cite{Omega} rather than taking	the value of 1:1 which
assumes exact isospin symmetry. The resultant change in $\cal{R}_{\rm 3}$ is 2.3\%, which is taken as the systematic uncertainty.

Assuming all the sources are independent and adding them in quadrature,
the total systematic uncertainties are obtained. All the systematical
uncertainties are summarized in Table~\ref{tab:errsum}.

\begin{table}[htbp]
\caption{\label{tab:errsum} Relative systematic uncertainties (\%) on the measurements of $\cal{R}$$_{1}$, $\cal{R}$$_{2}$, and $\cal{R}$$_{3}$.}
	\begin{tabular}{crrr}
		\hline\hline
		Sources & \multicolumn{1}{c}{$\cal{R}$$_{1}$} & \multicolumn{1}{c}{$\cal{R}$$_{2}$} & \multicolumn{1}{c}{$\cal{R}$$_{3}$} \\
	 	\hline
	    Detection-efficiency-related         &  -    &  -   &   2.2  \\
		MC statistics                        & 1.0   & 1.0  &   1.0  \\
		$\Omega(2012)$ resonance parameters  & 14.3  & 9.2  &   12.8 \\
		$\Xi^0$ mass                         & 4.2   &  -   &   3.2  \\
		Fit                                  & 10.4  & 9.9  &   7.8  \\
	    Ratio                                &  -    &  -   &   2.3  \\
	    Sum in quadrature                    & 18.2  & 13.6 &   15.7 \\
		\hline\hline
	\end{tabular}
\end{table}

In summary, using the entire data sample of 980~fb$^{-1}$ integrated luminosity
collected with the Belle detector, we search for the $\Omega(2012)^-$ resonance
in $\Omega_{c}^{0} \to \pip \Omega(2012)^- \to \pi^+ (\bar{K}\Xi)^{-}$. In
$\Omega_{c}^{0} \to \pip \Omega(2012)^- \to \pi^+ K^- \Xi^0$,
we find evidence for the $\Omega(2012)^-$ in the $K^- \Xi^0$ invariant mass spectrum
with a statistical significance of 4.0$\sigma$. In $\Omega_{c}^{0} \to \pip \Omega(2012)^- \to \pi^+ K_S^0 \Xi^-$,
a marginal $\Omega(2012)^-$ signal can be seen in the $K_S^0 \Xi^-$ invariant mass spectrum
with a statistical significance of 2.3$\sigma$. We perform a 2D simultaneous fit to
the two isospin decay modes, and the significance of $\Omega(2012)^-$ in
$\Omega_{c}^{0} \to \pi^+\Omega(2012)^- \to \pi^+ (\bar{K}\Xi)^{-}$ is 4.2$\sigma$,
including the systematic uncertainties. The ratios of the branching fractions
$\BR(\Omega_{c}^0 \to \pi^+ \Omega(2012)^-)\times \BR(\Omega(2012)^-
\to K^-  \Xi^0)/\BR(\Omega_{c}^0  \to \pi^+  K^-  \Xi^0)$, $\BR(\Omega_{c}^0 \to
\pi^+ \Omega(2012)^-) \times \BR(\Omega(2012)^- \to \bar{K}^0 \Xi^-)/\BR(\Omega_{c}^0 \to
\pi^+ \bar{K}^0 \Xi^-)$, and $\BR(\Omega_{c}^0 \to \pi^+ \Omega(2012)^-) \times \BR(\Omega(2012)^- \to (\bar{K} \Xi)^{-})/\BR(\Omega_{c}^0 \to \pi^+ \Omega^-)$ are measured to be (9.6 $\pm$ 3.2(stat.) $\pm$ 1.8(syst.))\%,
(5.5 $\pm$ 2.8(stat.) $\pm$ 0.7(syst.))\%, and 0.220 $\pm$ 0.059(stat.) $\pm$ 0.035(syst.), respectively.

We thank the KEKB group for the excellent operation of the
accelerator; the KEK cryogenics group for the efficient
operation of the solenoid; and the KEK computer group, and the Pacific Northwest National
Laboratory (PNNL) Environmental Molecular Sciences Laboratory (EMSL)
computing group for strong computing support; and the National
Institute of Informatics, and Science Information NETwork 5 (SINET5) for
valuable network support.  We acknowledge support from
the Ministry of Education, Culture, Sports, Science, and
Technology (MEXT) of Japan, the Japan Society for the
Promotion of Science (JSPS), and the Tau-Lepton Physics
Research Center of Nagoya University;
the Australian Research Council including grants
DP180102629, 
DP170102389, 
DP170102204, 
DP150103061, 
FT130100303; 
Austrian Science Fund (FWF);
the National Natural Science Foundation of China under Contracts
No.~11435013,  
No.~11475187,  
No.~11521505,  
No.~11575017,  
No.~11675166,  
No.~11705209;  
No.~11761141009;
No.~11975076;
No.~12042509;
Key Research Program of Frontier Sciences, Chinese Academy of Sciences (CAS), Grant No.~QYZDJ-SSW-SLH011; 
the  CAS Center for Excellence in Particle Physics (CCEPP); 
the Shanghai Pujiang Program under Grant No.~18PJ1401000;  
the Ministry of Education, Youth and Sports of the Czech
Republic under Contract No.~LTT17020;
the Carl Zeiss Foundation, the Deutsche Forschungsgemeinschaft, the
Excellence Cluster Universe, and the VolkswagenStiftung;
the Department of Science and Technology of India;
the Istituto Nazionale di Fisica Nucleare of Italy;
National Research Foundation (NRF) of Korea Grant
Nos.~2016R1\-D1A1B\-01010135, 2016R1\-D1A1B\-02012900, 2018R1\-A2B\-3003643,
2018R1\-A6A1A\-06024970, 2018R1\-D1A1B\-07047294, 2019K1\-A3A7A\-09033840,
2019R1\-I1A3A\-01058933;
Radiation Science Research Institute, Foreign Large-size Research Facility Application Supporting project, the Global Science Experimental Data Hub Center of the Korea Institute of Science and Technology Information and KREONET/GLORIAD;
the Polish Ministry of Science and Higher Education and
the National Science Center;
the Ministry of Science and Higher Education of the Russian Federation, Agreement 14.W03.31.0026; 
the Slovenian Research Agency;
Ikerbasque, Basque Foundation for Science, Spain;
the Swiss National Science Foundation;
the Ministry of Education and the Ministry of Science and Technology of Taiwan;
and the United States Department of Energy and the National Science Foundation.

\end{document}